\documentstyle[12pt,epsf]{article}

\begin{document}

\def\CA{{\cal A}}
\def\CB{{\cal B}}
\def\CC{{\cal C}}
\def\CD{{\cal D}}
\def\CE{{\cal E}}
\def\CF{{\cal F}}
\def\CG{{\cal G}}
\def\CH{{\cal H}}
\def\CI{{\cal I}}
\def\CJ{{\cal J}}
\def\CK{{\cal K}}
\def\CL{{\cal L}}
\def\CM{{\cal M}}
\def\CN{{\cal N}}
\def\CO{{\cal O}}
\def\CP{{\cal P}}
\def\CQ{{\cal Q}}
\def\CR{{\cal R}}
\def\CS{{\cal S}}
\def\CT{{\cal T}}
\def\CU{{\cal U}}
\def\CV{{\cal V}}
\def\CW{{\cal W}}
\def\CX{{\cal X}}
\def\CY{{\cal Y}}
\def\CZ{{\cal Z}}

\newcommand{\todo}[1]{{\em \small {#1}}\marginpar{$\Longleftarrow$}}
\newcommand{\ads}[1]{{\rm AdS}_{#1}}
\newcommand{\SL}[0]{{\rm SL}(2,\IR)}
\newcommand{\cosm}[0]{R}
\newcommand{\tL}[0]{\bar{L}}
\newcommand{\hdim}[0]{\bar{h}}
\newcommand{\bw}[0]{\bar{w}}
\newcommand{\bz}[0]{\bar{z}}
\newcommand{\be}{\begin{equation}}
\newcommand{\ee}{\end{equation}}
\newcommand{\lp}{\lambda_+}
\newcommand{\bx}{ {\bf x}}
\newcommand{\bk}{{\bf k}}
\newcommand{\tp}{\tilde{\phi}}
\hyphenation{Min-kow-ski}

\def\pref#1{(\ref{#1})}

\def\ie{{\it i.e.}}
\def\eg{{\it e.g.}}
\def\cf{{\it c.f.}}
\def\etal{{\it et.al.}}
\def\etc{{\it etc.}}

\def\adv{{\it Adv. Phys.}}
\def\ap{{\it Ann. Phys, NY}}
\def\cqg{{\it Class. Quant. Grav.}}
\def\cmp{{\it Comm. Math. Phys.}}
\def\jetp{{\it Sov. Phys. JETP}}
\def\jetpl{{\it JETP Lett.}}
\def\jp{{\it J. Phys.}}
\def\ijmp{{\it Int. J. Mod. Phys. }}
\def\inc{{\it Nuovo Cimento}}
\def\np{{\it Nucl. Phys.}}
\def\mpl{{\it Mod. Phys. Lett.}}
\def\pl{{\it Phys. Lett.}}
\def\pr{{\it Phys. Rev.}}
\def\prl{{\it Phys. Rev. Lett.}}
\def\prcl{{\it Proc. Roy. Soc.} (London)}
\def\rmp{{\it Rev. Mod. Phys.}}
\def\dash{-----------------    }

\def\p{\partial}

\def\apr{\alpha'}
\def\str{{str}}
\def\lstr{\ell_\str}
\def\gstr{g_\str}
\def\Mstr{M_\str}
\def\lpl{\ell_{pl}}
\def\Mpl{M_{pl}}
\def\varep{\varepsilon}
\def\del{\nabla}
\def\tr{\hbox{tr}}
\def\perp{\bot}
\def\half{\frac{1}{2}}
\def\p{\partial}

\def\IB{\relax\hbox{$\inbar\kern-.3em{\rm B}$}}
\def\IC{\relax\hbox{$\inbar\kern-.3em{\rm C}$}}
\def\ID{\relax\hbox{$\inbar\kern-.3em{\rm D}$}}
\def\IE{\relax\hbox{$\inbar\kern-.3em{\rm E}$}}
\def\IF{\relax\hbox{$\inbar\kern-.3em{\rm F}$}}
\def\IG{\relax\hbox{$\inbar\kern-.3em{\rm G}$}}
\def\IGa{\relax\hbox{${\rm I}\kern-.18em\Gamma$}}
\def\IH{\relax{\rm I\kern-.18em H}}
\def\IK{\relax{\rm I\kern-.18em K}}
\def\IL{\relax{\rm I\kern-.18em L}}
\def\IP{\relax{\rm I\kern-.18em P}}
\def\IR{\relax{\rm I\kern-.18em R}}
\def\IZ{\relax{\rm Z\kern-.5em Z}}

\renewcommand{\thepage}{\arabic{page}}
\setcounter{page}{1}
\newcommand{\nc}{\newcommand}
\nc{\beq}{\begin{equation}} \nc{\eeq}{\end{equation}}
\nc{\beqa}{\begin{eqnarray}} \nc{\eeqa}{\end{eqnarray}}
\nc{\lsim}{\begin{array}{c}\,\sim\vspace{-21pt}\\< \end{array}}
\nc{\gsim}{\begin{array}{c}\sim\vspace{-21pt}\\> \end{array}}

\newcommand{\drawsquare}[2]{\hbox{%
\rule{#2pt}{#1pt}\hskip-#2pt
\rule{#1pt}{#2pt}\hskip-#1pt
\rule[#1pt]{#1pt}{#2pt}}\rule[#1pt]{#2pt}{#2pt}\hskip-#2pt
\rule{#2pt}{#1pt}}

\newcommand{\Yfund}{\raisebox{-.5pt}{\drawsquare{6.5}{0.4}}}
\newcommand{\Ysymm}{\raisebox{-.5pt}{\drawsquare{6.5}{0.4}}\hskip-0.4pt%
        \raisebox{-.5pt}{\drawsquare{6.5}{0.4}}}
\newcommand{\Yasymm}{\raisebox{-3.5pt}{\drawsquare{6.5}{0.4}}\hskip-6.9pt%
        \raisebox{3pt}{\drawsquare{6.5}{0.4}}}
\newcommand{\Ythree}{\raisebox{-3.5pt}{\drawsquare{6.5}{0.4}}\hskip-6.9pt%
        \raisebox{3pt}{\drawsquare{6.5}{0.4}}\hskip-6.9pt
        \raisebox{9.5pt}{\drawsquare{6.5}{0.4}}}

\rightline{HUTP-98/A057, CALT68-2189}
\rightline{Fermilab-Pub-98/240-T, hep-th/9808017}
\vskip 1cm
\centerline{\Large \bf Holographic Probes }
\centerline{\Large \bf of Anti-de Sitter Spacetimes}
%
\vskip 1cm

\renewcommand{\thefootnote}{\fnsymbol{footnote}}
\centerline{{\bf Vijay
Balasubramanian${}^{1}$\footnote{vijayb@curie.harvard.edu}, 
Per Kraus${}^{2}$\footnote{perkraus@theory.caltech.edu},}} 
\centerline{{\bf Albion
Lawrence${}^{1}$\footnote{lawrence@string.harvard.edu}
 and Sandip P. Trivedi${}^{3}$\footnote{trivedi@fnth23.fnal.gov}}}
\vskip .5cm
\centerline{${}^1$\it Lyman Laboratory of Physics, Harvard University}
\centerline{\it Cambridge, MA 02138, USA}
\vskip .5cm
\centerline{${}^2$ \it California Institute of Technology}
\centerline{\it Pasadena, CA 91125, USA}
\vskip .5cm
\centerline{${}^3$ \it Fermi National Accelerator Laboratory}
\centerline{\it P.O. Box 500}
\centerline{\it Batavia IL, 60510, USA}

\setcounter{footnote}{0}
\renewcommand{\thefootnote}{\arabic{footnote}}

\begin{abstract}
We describe probes of anti-de Sitter spacetimes in terms of conformal
field theories on the AdS boundary.  Our basic tool is a formula that
relates bulk and boundary states -- classical bulk field configurations
are dual to expectation values of operators on the boundary. At the
quantum level we relate the operator expansions of bulk and boundary
fields.  Using our methods, we discuss the CFT description of local
bulk probes including normalizable wavepackets, fundamental and
D-strings, and D-instantons.  Radial motions of probes in the bulk
spacetime are related to motions in scale on the boundary,
demonstrating a scale-radius duality.  We discuss the implications of
these results for the holographic description of black hole horizons
in the boundary field theory.
\end{abstract}

\def\figloc#1#2{\bigskip\vbox{{\epsfxsize=4.5in \nopagebreak[3]
        \centerline{\epsfbox{diag#1.ps}} \nopagebreak[3] \centerline{Figure
        #1} \nopagebreak[3] {\raggedright\it \vbox{ #2 }}}} \bigskip }

\section{Introduction}

A recurring theme of recent work is that gravitational theories can
sometimes be formulated as gauge theories in fewer dimensions.  This
point of view has had some encouraging successes, but we still do not
understand how the famous problems of quantum gravity -- for example,
information loss in black hole evaporation -- are solved.  All of our
intuitions about gravity and spacetime physics are based on a
classical, geometric picture valid when $\hbar$ is small and the field
configurations macroscopic.  In this regime the spacetime physics
displays at least approximate locality and causality, and a
well-defined geometry in which free particles follow geodesics.  These
properties seem obscure in the gauge theory formulation.  Once we
understand their origin, we can investigate precisely when and how
they break down.

An important avenue for understanding these issues is the most recent
manifestation of the gravity-gauge theory connection: the
conjecture~\cite{juanads} that string theory on an anti-de Sitter
background is dual to a conformal field theory living on the
spacetime boundary.  So far, the proposal has been checked by
comparing spectra and low-order correlation functions of the dual
theories.  Such checks are based on a remarkably compact and powerful
statement of the equality between certain path integrals in the dual
theories~\cite{gkp,holowit}.  We would like to use this equality to
learn how the classical geometric description of the bulk emerges and
ultimately breaks down in the holographic boundary representation.

This article begins such a study by describing a variety of spacetime
probes from the boundary perspective.  Our basic tool is a compact
formula relating bulk and boundary states; specifically, the
asymptotic behavior of fluctuating, classical bulk fields is related
to expectation values of the dual boundary operators in excited
boundary states.  At the operator level we relate the quantized mode
expansion of bulk fields to a mode expansion of boundary operators, so
bulk quanta are dual to CFT states created by modes of boundary
operators acting on the vacuum.  We develop this formalism in
Sec.~\ref{sec:relbb} and, more carefully, in Appendix~\ref{app:corrs}.
This development continues the work in~\cite{us} which identified
fluctuating supergravity modes as dual to boundary states, and
non-fluctuating modes implementing boundary conditions as dual
to boundary sources.  Here, we will be interested mainly in the classical
limit of the fluctuating states and in bulk configurations generated
by brane sources.

Using the methods of Sec.~\ref{sec:relbb}, we discuss the
boundary description of three kinds of probes: D-instantons, F- and
D-strings, and dilaton wavepackets.  In all cases the characteristic
radial position of the bulk probe is mapped to the characteristic
scale of the boundary configuration, as understood on general grounds
by comparing the action of bulk isometries with conformal
transformations on the boundary~\cite{juanads}.  This scale-radius
duality gives rise to pleasantly physical interpretations of bulk
dynamics.  For instance, strings or particles move in AdS spacetime to
reduce gravitational potential energy; this is dual in the boundary
theory to the spreading of localized field distributions to reduce
gradient energy.  In Euclidean $\ads{5}\times S^5$ it has been
conjectured that a D-instanton at radial position $z$ is dual to an
instanton in $d=4$ SYM with scale size $z$~\cite{banksgreen, inst1,
kogan,greenetal}.  As an application of our methods we {\em derive} this
correspondence from the fundamental formulation of the AdS/CFT
conjecture given in~\cite{gkp,holowit}.  We conclude by discussing the
meaning of the classical limit on both sides of the AdS/CFT duality,
and by discussing the implications of our results for the holographic
representation of black hole horizons in the boundary gauge theory.

\section{Relating bulk and boundary states}
\label{sec:relbb}
In order to study how bulk geometry is encoded in the boundary
description, we wish to introduce probes into the AdS spacetime.  In
this section we develop methods that identify boundary configurations
corresponding to these probes.  Our basic technique is to study the
response of the bulk probe to a small change in boundary conditions.
The formulation of the AdS/CFT correspondence in~\cite{gkp,holowit}
then provides the expectation values of operators in the corresponding
boundary states.  This will also allow us to arrive at a quantized
formulation of the bulk theory from the boundary perspective.  In this
section we will ignore some subtleties~\cite{freedcorr} which are
unimportant for most of the considerations of this paper.  A more
careful treatment is provided in Appendix~\ref{app:corrs}.

We work with Poincar\'e coordinates for anti-de Sitter space.  The
metric of $\ads{d+1}$ in these coordinates is:
\begin{equation}
ds^2 = {\cosm^2 \over z^2}( -dt^2 + d\vec{x}_{d-1}^2 + dz^2 )
\label{eq:poinmet}
\end{equation}
(We set $\cosm=1$ in this section.)  Poincar\'e coordinates only cover
a patch of the global spacetime, and $z=0$ is the boundary of AdS
while $z=\infty$ is the horizon.  (See~\cite{us} for more
details and Penrose diagrams.)  Euclidean AdS is obtained by taking
$t$ to $it$.

\subsection{Euclidean signature}
\label{sec:eucsig}
The AdS/CFT correspondence is formulated in~\cite{gkp,holowit} as:
\begin{equation}
Z(\phi_i)=e^{-S(\phi_i)}=\langle e^{\int_B \phi_{0,i}\CO^i} \rangle
\label{eq:bbrel}
\end{equation}
where $S(\phi_i)$ is the effective action as a function of the bulk
field $\phi_i$, $\phi_{0,i}$ is the boundary value of $\phi_i$ (up to
a scaling with the radial coordinate) , and
$\CO^i$ is the dual operator in the CFT.  The expectation value on the
right is evaluated in the CFT vacuum.  We can read (\ref{eq:bbrel}) as
saying that boundary conditions for the bulk theory are dual to
sources in the boundary theory.  In other words, field theory in
Euclidean AdS, expanded around a background approaching $\phi_{0,i}$ at the
boundary, is described by a CFT deformed by the addition of a source.
By functionally differentiating  we find that:
\begin{equation}
\frac{\delta}{\delta \phi_{0,i}(\bx)}\, (-S(\phi_i))
 =
\langle \CO^i(\bx) \rangle_{\phi_{0,i}}
\label{eq:euc1pt}
\end{equation}
where the subscript $\phi_{0,i}$ indicates that the expectation value on the
right hand side is
computed in the presence of the source term $\int \phi_{0,i}\CO^i$.
We will use this relation to learn about the expectation values of
boundary operators in the presence of bulk probes.

\paragraph{Massive scalar: } As an example, let us study a massive
scalar field with a quadratic bulk action:
\begin{equation}
S(\phi) = \frac{1}{2} \int \! d^{d+1}x \, \sqrt{g} 
\left( |d\phi |^2+m^2 \phi^2 \right),
\label{action}
\end{equation}
Consider a solution to the bulk equations of motion that approaches
$\phi_0(x)$ at the boundary (up to a scaling with the radial
coordinate $z$).  The unique solution regular in the bulk behaves as
$\phi(z,\bx) = z^{2h_-} \phi_0(\bx)$ as $z \rightarrow 0$ and can be
written as:
\begin{equation}
\phi(z,\bx) = c \int \! d^d\bx' \, \frac{z^{2h_+}}{\left(z^2+|\bx-\bx'|^2
\right)^{2h_+}}\,\phi_0(\bx').
\label{sol}
\end{equation}
where we have used the bulk-boundary propagator~\cite{holowit} and
\begin{equation}
h_\pm = {d \over 4} \pm {\sqrt{d^2 + 4 m^2} \over 4}
\equiv {d \over 4} \pm {\nu \over 2}
\label{eq:hpm}
\end{equation}
The presence of this classical configuration corresponds to the addition of a 
source
$\int \phi_0 \CO$ to the boundary theory.

We now  apply (\ref{eq:euc1pt}) by computing the functional
derivative on the left hand side.  To do so we perturb
around $\phi(z,\bx)$ by a small fluctuation $\delta \phi$ and evaluate
the resulting change in the action.  After integrating (\ref{action})
by parts and using the equations of motion, the variation becomes a
surface term at the boundary:
\begin{equation}
\delta S(\phi) = \int_B d\Sigma^{\mu} \, \p_{\mu} \phi \, \delta\phi.
\label{eq:surf}
\end{equation}
Evaluating this quantity near the boundary at $z=0$ is
delicate~\cite{freedcorr}.  In this section we will
follow~\cite{holowit} by considering contributions to the integrand
of~(\ref{sol}) from the region $|\bx - \bx'|\neq 0$.  This procedure
amounts to ignoring certain contact terms and normalization issues as
we discuss in Appendix~\ref{app:corrs} and gives, as $z \rightarrow
0$:
\begin{equation}
\frac{\p \phi}{\p z} = c \, (2h_+)\, z^{2h_+-1} \int \! d^d \bx'\,
\frac{\phi_0(\bx')}{|\bx-\bx'|^{2(2h_+)}}.
\end{equation}
The  perturbation is written as $\delta \phi = z^{2h_-} \delta \phi_0$, and
we use $d\Sigma^{0} = z^{1-d} d^d\bx'$.  Then
\begin{equation}
\delta S(\phi) = c\, (2h_+) \int \! d^d\bx \,\,d^d\bx'\, 
\frac{\phi_0(\bx') \, \delta \phi_0(\bx)}{|\bx-\bx'|^{2(2h_+)}}.
\end{equation}
Using the relation (\ref{eq:euc1pt}) we derive
\begin{equation}
\langle \CO(\bx) \rangle_{\phi_0} = - c(2h_+)
\int \! d^d\bx'\, \frac{\phi_0(\bx')}{|\bx-\bx'|^{2(2h_+)}}.
\label{Oexp}
\end{equation}

We have learned that in the presence of the source term
$\int \phi_0 \CO$ the operator $\CO$ has acquired an  expectation value
given by the right hand side of (\ref{Oexp}). This matches what we expect
from the CFT by direct calculation:
\begin{equation}
\langle \CO(\bx) \rangle_{\phi_0} = \langle \CO(\bx)e^{\int\!\phi_0 \CO} \rangle
\approx \int \!d^d\bx' \, \phi_0(\bx')\langle \CO(\bx)\CO(\bx')\rangle
\approx
-\int \! d^d\bx' \, \frac{\phi_0(\bx')}{|\bx-\bx'|^{2(2h_+)}},
\end{equation}
where the form of the two point correlator follows from scale invariance.

\paragraph{Interpretation: }
The AdS/CFT correspondence in (\ref{eq:bbrel}) states that turning on a
bulk mode which behaves as $z^{2h_-}\phi_0(\bx)$ near the boundary
is dual to including a source term $\int\!\phi_0 \CO$ in the CFT.  As
discussed in~\cite{us}, the growth of such modes near the
boundary indicates that they are  non-fluctuating
classical backgrounds.  In effect, the presence of the mode
redefines the Hamiltonian of the theory, since fluctuations should
take place on top of this background.  This is mirrored in 
the CFT by a modification of the action by the addition of a source
term.

In the bulk, a mode with leading boundary behaviour $z^{2h_-}\phi_0$
induces a {\em subleading} component behaving as $z^{2h_+}\tp$.
(This is seen by expanding (\ref{sol}) in powers of $z$.)  The
corresponding statement in the CFT is that the addition of the source
induces an expectation value for $\CO$.  In fact, our analysis showed
that $\langle\CO(\bx)\rangle_{\phi_0} \sim \tp(\bx)$, so that operator
expectation values and bulk field components behaving as $z^{2h_+}$
are precisely dual.  This duality is the prevailing theme of the
present work.

\paragraph{Bulk sources: }  In the above example, the solution
$\phi(z, \bx)$ was completely determined by the boundary value $\phi_0$
and the requirement of regularity in the bulk.  However, as we shall 
see in Sec.~\ref{sec:dinst}, this uniqueness fails when we admit
singular fields corresponding to sources in the bulk.  Such bulk
sources contribute subleading pieces to the fields at the boundary
which modify (\ref{sol}) and contribute to operator expectation
values.  So once again we will find that subleading pieces of the bulk
fields are dual to boundary expectation values.  In this way, we will
show in Sec.~\ref{sec:dinst} that (\ref{eq:euc1pt}) implies that a
bulk D-instanton in $\ads{5}$ is dual to an instanton in the boundary
Yang-Mills theory.

\subsection{Lorentzian signature}
\label{sec:lorsig}
The crucial new feature of Lorentzian signature is that the bulk wave
equation admits propagating, normalizable mode solutions.  Such modes
describe the physical, low energy excitations of the spacetime; their
explicit forms have been worked out in (\cite{avisetal, brfreed, mez,
us}) and they behave as $z^{2h_+}$ near the boundary.  These 
normalizable modes form the Hilbert space of the bulk theory.  The
possible boundary conditions for fields in AdS spacetime are
encoded by the choice of non-normalizable mode solutions behaving as
$z^{2h_-}$ near the boundary.  As argued in~\cite{us}, the
normalizable and non-normalizable solutions are dual to states and
sources respectively in the boundary conformal field theory.  Here we
make explicit the map between bulk and boundary states.  So given a
bulk field $\phi_i$ approaching $z^{2h-}\phi_{0,i}$ at the boundary,
we write the Lorentzian bulk-boundary correspondence as:
\begin{equation}
Z(\phi_i)=e^{iS(\phi_i)}=\langle s| e^{i\int_B \phi_{0,i}\CO^i} |s\rangle
\label{eq:lorrel}
\end{equation}
Here $|s\rangle$ represents the CFT state that is dual to the bulk
state.  Operator expectation values in excited CFT states will differ
from their vacuum values; we write:
\begin{equation}
\frac{\delta}{\delta \phi_{0,i}(\bx)}\, S(\phi_i)
 =
\langle s| \CO^i(\bx) | s \rangle_{\phi_{0,i}}
\label{eq:lor1pt}
\end{equation}
where the subscript indicates that the expectation is computed in the
presence of a source term.   We will also show that this can be turned
into a statement relating quantized field operators in the bulk to operators 
in the boundary.

\paragraph{Massive scalar: }   As an example, we return again to the
free massive scalar with action (\ref{action}).  Now a general
nonsingular solution of  the bulk equations approaching
$z^{2h_-}\phi_0(\bx)$ at the boundary can be written as:
\begin{equation}
\phi(z,\bx) = \phi_n(z,\bx)+c \int \! d^d\bx'\,\frac
{z^{2h_+}}{\left(z^2+|\bx-\bx'|^2
\right)^{2h_+}}\,\phi_0(\bx')
\label{lorsol}
\end{equation}
where $\phi_n$ is a normalizable mode, and $|\bx-\bx'|^2 = -(t-t')^2 +
\sum_{i=1}^{d-1}(x_i-x_i')^2$.   Here we have used a bulk-boundary
propagator obtained by continuation from Euclidean space 
in~\cite{holowit}.  There are ambiguities in this choice whose meaning
is discussed in Sec.~\ref{sec:operator} and Appendix~\ref{app:corrs}.
We now repeat the procedure used in
Euclidean signature, dropping contact terms as before, and paying
attention to the extra contribution from $\phi_n$.  (There are
subtleties in this procedure - see Appendix~\ref{app:corrs}.) Using
$\phi_n(z,\bx) \rightarrow z^{2h_+}
\tp_n(\bx)$ as $z \rightarrow 0$, we find:
\begin{equation}
\langle \tp_n| \CO(\bx) | \tp_n \rangle_{\phi_0} = 
(2h_+) \tp_n(\bx) + c(2h_+) \int \! d^d\bx' \,
\frac{\phi_0(\bx')}{|\bx-\bx'|^{2(2h_+)}}
\label{lorexp}
\end{equation}
where we have indicated that the CFT is in the excited state $|\tp_n
\rangle$.\footnote{The normalizations produced by this naive
Lorentzian treatment are not correct - see Appendix~\ref{app:corrs}.}
So $\CO$ gets an expectation value from two distinct contributions:
from the excited state and from the source that has been turned on.
Note that $|\tp_n
\rangle$ is a ``coherent'' state on the boundary in which operators
have non-vanishing expectation values.

\paragraph{Interpretation: } In Lorentzian AdS, normalizable
 and non-normalizable modes are dual to states and sources
respectively.  As we have seen, operator expectation values are
affected by both the state and the source.  Nevertheless, as in the
Euclidean case, the component of the total bulk field behaving as
$z^{2h-}$ defines the source while the component behaving as
$z^{2h_+}$ gives rise to the boundary expectation value.  We are free
to set $\phi_0$ to zero if we wish, so that the sources are turned off
-- then we are studying states of the original unmodified CFT.

\paragraph{Bulk sources: } In the example above, we considered
linearized wave equations for a massive scalar.  It is possible to
consider fully non-linear solutions with possible bulk singularities
due to sources.  For instance, the field configuration arising from a
D-brane in the AdS geometry would be of this type.  The treatment
based on (\ref{eq:lor1pt}) is equally valid in this case -- linearized
fluctuations around the fully non-linear solution lead by integration
by parts to the same surface integrals. When $\phi_0$ vanishes
(\ref{lorexp}) will also continue to hold, but in general nonvanishing
$\phi_0$ will push the field configuration into the non-linear regime
and bulk interactions will become important.  Equipped with
(\ref{lorexp}), in the next sections we will determine the dual
boundary description of various solitonic objects in the bulk.

\subsection{Operator formulation}
\label{sec:operator}
We would also like a more microscopic mapping at the level of
individual quantum states.  This is obtained by regarding fields as
quantized operators.  In particular, we may write $\phi_n$ in terms of
a mode expansion:
\begin{equation}
\hat{\phi}_n = \sum_k \left[a_{k}\phi_{n,k} + a^{\dagger}_{k} \phi^*_{n,k}
\right]
\label{phiop}
\end{equation}
and similarly for the boundary operator $\CO$,
\begin{equation}
\hat{\CO} = \sum_k \left[b_{k}\tp_{n,k} + b^{\dagger}_{k} \tp^*_{n,k}
\right],
\label{Oop}
\end{equation}
where as before, $\tilde{\phi}(\bx)$ is the boundary value of the
component of the bulk mode $\phi(z,\bx)$ scaling as $z^{2h_+}$ as $z
\rightarrow 0$.   Note that the modes appearing in the expansion of
$\hat{\phi}_n$ satisfy an on-shell condition $ (\Box
-m^2)\phi_{n,k}=0$, whereas the modes $\tp_{n,k}$ do not satisfy any
wave equation on the boundary, but are instead a fully complete set of
functions.  Interpreting (\ref{lorexp}) as an operator
statement we conclude that $a_k = b_k$, and that $a^{\dagger}_{k}|0
\rangle = |k\rangle$ where $|k\rangle$ is a ``one
particle'' state created by a single application of $b_k^\dagger$.
(It is intriguing that creation and annihilation operators of
elementary bulk fields are related to composite operators on the
boundary, leading us to identify $\sum_k b^\dagger_{k} b_{k}$ as a
particle number operator.)  In other words, bulk states described by
quanta occupying normalizable modes are dual to CFT states described
by acting on the vacuum with modes of the appropriate boundary
operator.  This provides a direct correspondence between bulk and CFT
states.\footnote{Related issues are discussed in~\cite{tometal}.}

\paragraph{Choice of vacuum: } Quantum field theory in curved
space can usually accomodate a variety of inequivalent vacua,
corresponding to different definitions of positive frequency.  In the
present context, the choice of vacuum affects the formalism in two
places. First, mode solutions in (\ref{phiop},\ref{Oop}) have 
positive frequency with respect to a particular time coordinate, here
taken to be Poincar\'e time.  A different choice of time can lead to
an inequivalent vacuum state, related to the original vacuum by a
Bogoliubov transformation.  Second, we have made a particular choice
for the form of the bulk-boundary propagator in (\ref{lorsol}) which
we obtained by continuation from Euclidean space.  This is the
appropriate propagator to use when perturbing around the Poincar\'e
vacuum.  However, alternative vacua can be chosen by changing the mode
expansions (\ref{phiop},\ref{Oop}) and modifying the bulk-boundary
propagator.  The latter modification will involve adding a
normalizable mode to the original propagator, which leaves unchanged
its $z^{2h_-}$ dependence near $z=0$.  These issues are discussed in
greater detail in Appendix~\ref{app:corrs}

\subsection{Radial isometry and boundary scale transformation}
In subsequent sections we will use the formalism developed above to
study the boundary representation of probes in the bulk.  A recurring
theme will be a duality between characteristic radial positions in the
bulk and characteristic scales in the boundary.  Let us review how
this arises.   The Poincar\'e metric (\ref{eq:poinmet}) has a
radial isometry:
\begin{equation}
(\vec{x},t) = \bx \rightarrow \lambda \bx ~~~~~~~;~~~~~~~
z \rightarrow \lambda z
\label{eq:isom}
\end{equation}
As we have discussed, boundary expectation values for a massive scalar
are dual to the component of the bulk field scaling as $z^{2h_+}
\tilde{\phi}(\bx)$ near the boundary.  So consider a one
parameter family of bulk solutions of the form:
\begin{equation}
\phi^\lambda(z,\bx) = \phi(\lambda z, \lambda \bx)
\label{eq:bulkisom}
\end{equation}
According to (\ref{lorexp}), the expectation value of the boundary
operator in the corresponding state behaves as:
\begin{equation}
\langle \phi^\lambda | \CO(\bx) | \phi^\lambda \rangle
= \lambda^{2h_+} \langle \CO(\lambda\bx) \rangle
\label{eq:boundscale}
\end{equation}
So the radial isometry generates scale transformations on the
boundary.  In the examples studied below we will explicitly see that
the boundary configuration spreads out as the bulk probe falls towards
the Poincar\'e horizon.

\section{Instanton probes}
\label{sec:dinst}

Our first example of a bulk probe is a D-instanton in $\ads{5}\times
S^5$.  The methods developed in the previous section will show that it
is dual to a boundary Yang-Mills instanton.  A particular consequence
is a duality between the radial position of the bulk object and the
scale size on the boundary.  D-instanton solutions in AdS space have
been discussed in~\cite{inst1,kogan,greenetal} where the close
similarity between the bulk dilaton profile and the boundary instanton
was noted.  Our main point in this section is that this fact
follows from the general considerations of Sec.~\ref{sec:relbb}.  This
{\em derives} the duality between the D-instanton and the Yang-Mills
instanton from the fundamental formulation of the AdS/CFT
correspondence in (\ref{eq:bbrel}).

According to the methods of Sec.~\ref{sec:relbb}, to determine the
boundary expectation values corresponding to a D-instanton we require
the form of the bulk solution near the boundary.  We will use
Poincar\'e coordinates for $\ads{5}$, so that the metric is given by
(\ref{eq:poinmet}) with $R^4= 4 \pi g_s N \alpha'^2$.  D-instanton
solutions have been presented in~\cite{inst1,kogan,greenetal}; only
the dilaton ($\phi$) and the axion ($\chi$) are turned on, while the
$\ads{5}\times S^5$ Einstein metric is unchanged.  Dimensional
reduction of these fields on $S^5$ produces a Kaluza-Klein tower of
modes on $\ads{5}$.  Here we are only interested in the behaviour as
$z\rightarrow 0$ of the massless five dimensional dilaton and axion
that couple to $Tr(F^2)$ and $Tr(F\tilde{F})$ on the
boundary.\footnote{These couplings were first studied
in~\cite{klebetal}.}  This is given by:\footnote{There is some
disagreement between~\cite{inst1,kogan} and~\cite{greenetal}
concerning whether the the D-instanton should be localized on the
$S^5$.  The difference will lie in the excitation of the Kaluza-Klein
harmonics that are massive fields on $\ads{5}$.  Including these
excitations which fall off faster at the boundary will give
expectation values to dual higher dimension operators that were
identified in~\cite{holowit}.  The authors
of~\cite{inst1,kogan,greenetal} all agree on the asymptotic form of
the massless $\ads{5}$ fields which is all that we require here.}
\begin{eqnarray}
e^{\phi} &=& g_s + c~{ z^4 ~\tilde{z}^4 \over [\tilde z^2 + \vert \vec x-
\vec x_a \vert^2]^4}  \cdots,
\label{soldilaton} \\
\chi&=& \chi_{\infty} \pm (e^{- \phi} - 1 /g_s ) 
\label{solaxion}
\end{eqnarray}
Here $\tilde{z}$ is the radial position of the D-instanton. 
The constant $c$ in (\ref{soldilaton}) 
can be determined by requiring that the D instanton
carries the correct axionic charge. That is:
\beq
\label{normc}
{1 \over 2 \kappa_{10}^2} \int e^{2 \phi}  \: \partial_{\mu} 
\chi  ~dS^{\mu} = 2 \pi,
\eeq
yielding
\beq
\label{valc}
c= {24 \pi \over N^2}. 
\eeq

\paragraph{Boundary expectations: } 
We can now use the methods of Sec.~\ref{sec:relbb} to derive 
the expectation values of boundary operators.  The 
five-dimensional dilaton action is:
\beq
\label{sdilaton}
S_{\phi} = -{1 \over 4 \kappa_{5}^2} \int d^{5}x \sqrt{g}g^{\mu \nu} 
\partial_{\mu} \phi \partial_{\nu} \phi + \cdots \cdot
\eeq 
where $1/\kappa_5^2 = V_5 R^5/\kappa_{10}^2$, $\kappa_{10}$ is the 10
dimensional Newton constant and $V_5 = \pi^3 $ is the volume of the
unit 5-sphere.  We start with the dilaton background for the D-instanton
(\ref{soldilaton}) and add a small perturbation $\delta \phi$.  The
resulting change in the action is a boundary term:
\beq
\label{dels}
\delta S= -{1 \over 2 \kappa_{5}^2} \int d^4x {R^3 \over z^3}
\delta \phi ~ \partial_z \phi
\eeq
It follows that the functional derivative with respect to the boundary
configuration $\phi_0$ is:
$(\delta S /  \delta \phi_0 ) = -(1 / 2 \kappa_{5}^2) ~R^3 ~
\partial_z \phi $.
Using $\phi$ in (\ref{soldilaton}) and the relation 
$R^4= \kappa_{10} N / 2 \pi^{5/2}$ gives:
\beq
\label{finaldels}
{\delta S \over \delta \phi_0(\vec x )}= -{48 \over 4 \pi g_s} {\tilde{z}^4 
\over [\tilde{z}^2 + \vert \vec x  - \vec x_a \vert^2]^4 }\ . 
\eeq
We learn from~\cite{holowit,gkp} that the $\ads{5}$ massless dilaton
couples to $Tr(F^2)$ in the boundary CFT.  Choosing the normalization
$S_{YM}= (1/ 4 g_{YM}^2) \int d^4x \, Tr(F^2) + \cdots$ for the
Yang-Mills action gives:
\beq
\label{delsym}
{\delta S_{YM} \over \delta \phi_0 (\vec x)}= -{1\over 4 g_{YM}^2} 
\langle Tr(F^2(\vec x)) \rangle  \ .
\eeq
Equating (\ref{finaldels}) and (\ref{delsym}) and using $4 \pi g_s=
 g_{YM}^2$, we find:
\beq
\label{exp}
{1 \over 4 g_{YM}^2} <Tr F^2(\vec x)> = {48 \over g_{YM}^2} {\tilde{z}^4 
\over [\tilde{z}^2 + \vert \vec x  - \vec x_a \vert^2]^4 }\ , 
\eeq
which is exactly the Yang-Mills field strength in an instanton
background.  So, as advertised, a D-instanton in the bulk
is precisely dual to a Yang-Mills instanton in the boundary theory.

This discussion of the dilaton can be extended to the axion which
yields the expected $\langle F \tilde{F} \rangle$ for an instanton.
Since the $\ads{5}$ metric is unchanged, we learn that the expectation
value of the stress tensor $\langle T_{\mu\nu} \rangle$ vanishes in an
instanton background.  This is easily checked; the stress tensor is:
\be
T_{\mu\nu} =	\frac{1}{4} g_{\mu\nu} \, Tr F^{\rho\sigma} F_{\rho\sigma}
	- Tr F_\mu^\rho F_{\rho\nu}\ .
\ee
The two terms cancel for the Yang-Mills instanton.  Similarly, the
NS-NS B-field in an $S$-wave on $S^5$ is known to be dual to a
dimension 6 operator in the Yang-Mills theory~\cite{arvind}.  This
operator was derived in~\cite{dastriv,ferrara}:
\be
	\CO^{(6)}_{\mu\nu} \propto
	\tr \left[ \frac{1}{2} F_{\left[\nu\right.\alpha}
		F^{\alpha\beta}F_{\beta\left.\mu\right]}
	+ \frac{1}{8} F_{\alpha\beta}F^{\alpha\beta}
		F_{\mu\nu} \right]\ ,
\ee
where we have antisymmetrized the indices $\mu$~,~$\nu$.  Again, it is
simple to check that the first term vanishes upon antisymmetrization
and the second term vanishes identically.\footnote{Note, however, that
the symmetrized part of the first term does not vanish.  In general we
expect that there will be all sorts of combinations of $F$ that will
not vanish in this background.  This is not a surprise -- for example,
in the presence of a classical bulk configuration, we expect that many
operators corresponding to multi-particle bulk states will have
expectation values.}  Finally, the D-instanton action $S_{Dinst} =
2\pi/g_s$ coincides with the YM instanton action $S_{YM} =
8\pi^2/g^2_{YM}$ using $4\pi g_s = g^2_{YM}$.

\paragraph{Scale-radius duality and bulk locality: }
The above duality between the D-instanton and YM instanton provides
the first example of a phenomenon we will call scale-radius duality.
The characteristic radial position of the D-instanton is $\tilde{z}$.
On the boundary $\tilde{z}$ is the characteristic scale of the YM
instanton.   A D-instanton closer to the horizon at $\tilde{z} =
\infty$ is mapped into a fatter boundary object~\cite{banksgreen}.   
More specifically, the action of the isometries (\ref{eq:isom})
translates the D-instanton both radially and parallel to the boundary.
The corresponding conformal transformation of the boundary instanton
rescales it and translates it at the same time.  

This relation has interesting consequences for the emergence of local
physics in the bulk when we apply our techniques to multi-instanton
solutions.  Consider two D-instantons at very different radial
positions $z$, but at the same coordinate $\vec{x}$ parallel to the
boundary.  These are dual to two coincident YM instantons with widely
different scale sizes.  Locality of the bulk objects is expected at
large $N$ when classical physics is valid.  In this limit the
collective coordinates of the boundary configuration should
approximately decouple into two separate sets associated with
instantons of two different scale sizes (\ie\ the metric on the moduli
space is block-diagonal in this region).  Such behaviour typically
occurs for instantons at large spatial distances.  Here we learn that
a large difference in scale size will also cause a separation of
collective coordinates.  Turning this around, the approximate
non-interaction of collective coordinates of coincident instantons at
widely different scales translates into approximate locality of the
bulk physics.

We can also consider the interaction between instantons and
anti-instantons.  By evaluating the probe action of a D-instanton in
the background of the anti-D-instanton, we find a bulk interaction of
the form:
\beq
\label{perequation}
\delta S \sim  \frac{(z_1 z_2)^4}{(z_1-z_2)^8},
\eeq
where the D-instantons are at coincident $\vec{x}$ positions and
$z_{1,2}$ are their radial positions.  In deriving (\ref{perequation})
we have used the asymptotic form of the dilaton, which follows from
the required fall-off of the dilaton and $SO(1,5)$ invariance (it is
also consistent with \cite{inst1,kogan,greenetal}).  From
(\ref{perequation}) it follows that in the conformal large $N$
Yang-Mills theory we expect the interaction between coincident
instantons and anti-instantons to fall off as the eighth power of
difference in scale size.  Such behavior is not evident in
perturbative gauge theory, and presumably arises from the sum of
planar diagrams in the large $N$ limit.

\section{String probes}
\label{sec:stringpr}

Fundamental and D-strings are particularly interesting probes of
Lorentzian $\ads{5}$.  Working in Poincar\'e coordinates
(\ref{eq:poinmet}), we will find the Yang-Mills description of string
solitons stretched parallel to the AdS boundary, and placed at fixed
radial positions $\tilde{z}$.  Once again, the characteristic position
in the bulk will be mapped to the characteristic scale on the
boundary, and motion towards the Poincar\'e horizon appears as a
fattening of the boundary flux tube.  A string placed at a fixed
radial position is not a solution to the equations of motion since it
can reduce its potential energy by falling towards the horizon.
Nevertheless, since the analysis of a static string is technically
clearer, we imagine that it is stabilized by an external force. We
will find that the corresponding boundary flux could reduce its energy
by spreading and must be similarly stabilized.\footnote{Related
discussions appear in~\cite{tometal}.} This analysis is readily
generalized to slowly moving strings.

\paragraph{Strings as fluxes: }  First, we establish that fundamental (F)
and Dirichlet (D) strings in $\ads{5}$ are described by 
electric and magnetic fluxes in the boundary gauge theory.  This is
easily shown by starting with the worldvolume action for D3-branes
(with Higgs fields suppressed):
\begin{equation}
S_{D3}=-T_3 {\rm Tr}\int\! d^4\sigma\, \left\{e^{-\phi}
\sqrt{-\det(G_{mn} +2\pi \apr F_{mn}+B_{mn})}-\frac{1}{4}\epsilon^{mnpq}
C^{(2)}_{mn}F_{pq}\right\}
\end{equation}
(We have written the nonabelian Born-Infeld action appropriate to
commuting background fields~\cite{tseytlin}.)   The action
includes terms of the form: 
\begin{equation}
\int \!d^4\sigma \,B^{mn}{\rm Tr}F_{mn}  \quad\quad {\rm and} \quad\quad
\int \!d^4\sigma \, \epsilon^{mnpq}C^{(2)}_{mn}{\rm Tr}F_{mn},
\label{YMcouplings}
\end{equation}
where $B_{mn}$ and $C^{(2)}_{mn}$ are the NS-NS and RR 2-forms
respectively.  An F-string extended in the $x^i$ direction should
couple to $B_{ti}$, and so is described by a non-vanishing value of
$E_{i}={\rm Tr}F_{ti}$, in other words an electric flux.\footnote{The
source for the $U(1)$ electric flux in the Yang-Mills theory is not
associated with the dynamical part of $B_{mn}$, which couples to a
dimension 6 operator in the CFT~\cite{arvind,dastriv}, but rather with pure
gauge degrees of freedom which contribute to surface integrals for
conserved charges at infinity.   It may seem surprising that
(\ref{YMcouplings}) shows that bulk strings are related to the $U(1)$
part of the $U(N)$ boundary theory since this factor has been argued
to decouple~\cite{romansetal,holowit,oferwit}.  However, it is more
accurate to consider the $U(1)$  as ``frozen'' after inclusion of all
external probes and VEVs.  The remaining $SU(N)$ part of the theory is
dynamical and the rest of this section studies the expectation of
$Tr(F^2)$ with a trace in $SU(N)$.   We are grateful to O. Aharony for
correspondence regarding this issue.}
Similarly, the D-string corresponds to a magnetic flux
$B_{i}=\epsilon_{ijk}{\rm Tr}F_{jk}$.  This is consistent with
S-duality, which interchanges F and D strings in the bulk, and electric
and magnetic fields in the Yang-Mills theory.

\paragraph{Bulk strings: }  To describe bulk F and D strings we start 
with their worldsheet actions, which include the terms
\begin{eqnarray}
S_F&=&-\frac{1}{2\pi \apr}\int \!d^2\sigma \, \left\{ 
\sqrt{-\det g_{\mu\nu}\p_{m}X^\mu \p_nX^\nu}
-\frac{1}{2}\epsilon^{mn}B_{\mu\nu}\p_{m}X^{\mu}\p_{n}X^{\nu} \right\}
\label{Fstringaction}
\\ 
S_D&=&-\frac{1}{2\pi \apr g_s}\int \!d^2\sigma \, \left\{ e^{-\phi}
\sqrt{-\det g_{\mu\nu}\p_{m}X^\mu \p_nX^\nu}
-\frac{1}{2}\epsilon^{mn}C^{(2)}_{\mu\nu}\p_{m}X^{\mu}\p_{n}X^{\nu} \right\}
\label{Dstringaction}
 \end{eqnarray}
We have included couplings due to the string frame metric 
$g_{\mu\nu}$, the dilaton $\phi$, and the 2-forms $B_{\mu\nu}$, 
$C^{(2)}_{\mu\nu}$.  Now consider static strings extended in the
$x^1$ direction, parallel to the boundary.   In static gauge,
\begin{equation}
t(\sigma^m)=\sigma^0 \quad\quad z(\sigma^m)=\tilde{z}={\rm
constant} \quad \quad 
x^1(\sigma^m)=\sigma^1 \quad\quad \vec{x}_\perp(\sigma^m)=\vec{x}_{a\perp}.
\end{equation}
Here $\vec{x}_{a\perp}$ are directions orthogonal to the string but
parallel to the boundary.  Evaluating $S_F$, $S_D$ in the $\ads{5}$
background (\ref{eq:poinmet}) gives the potential energy of the static
strings (per unit coordinate length):
\begin{equation}
V_F =  \frac{g_{YM}}{2\pi}N^{1/2}\frac{1}{\tilde{z}^2} \quad\quad\quad
V_D = \frac{2}{g_{YM}}N^{1/2}\frac{1}{\tilde{z}^2},
\label{potentials}
\end{equation}
Here we used the relations $R^4= 4 \pi g_s N \alpha'^2$ and $4\pi g_s
= g_{YM}^2$ that are appropriate to $\ads{5}$.  The F and D string
potentials are related by S-duality: $g_{YM}^2/4\pi \rightarrow
(g_{YM}^2/4\pi)^{-1}$.

\paragraph{Boundary expectations: }
A bulk string is dual to a boundary CFT state in which various
operators have expectation values.  For example, the actions
(\ref{Fstringaction},\ref{Dstringaction}) will induce long range
fields for $B_{tx^1}$, $C^{(2)}_{tx^1}$ in the presence of F and D
strings respectively.  The results of Sec.~\ref{sec:relbb} and the
couplings (\ref{YMcouplings}) then yield nonvanishing values for
$\langle {\rm Tr}F_{tx^1}
\rangle_{F}$ and $\langle {\rm Tr}F_{x^2x^3} \rangle_{D}$,
corresponding to electric and magnetic fluxes.  Rather than evaluating
these explicitly, we focus on the expectation value for ${\rm Tr}F^2$,
which couples to the dilaton $\phi$.  Note that this trace is in the
$SU(N)$ part of the gauge group.  First we work out the long range
dilaton field produced by string sources via their linear coupling to
the dilaton.  Although it may appear from (\ref{Fstringaction}) that
the F-string does not couple to the dilaton, this is simply because
the actions are written in the string frame.  Working instead in the
Einstein frame, with $g^{E}_{\mu\nu}=(g_s e^{\phi})^{1/2}g_{\mu\nu}$,
we find the couplings:
\begin{equation}
S_F = -\frac{g_{YM}}{4\pi}N^{1/2} \int \! d^2 \sigma\, \frac{\phi}{\tilde{z}^2}
\quad\quad
S_D = \frac{1}{g_{YM}}N^{1/2} \int \! d^2\sigma\, \frac{\phi}{\tilde{z}^2}.
\label{dilcoupling}
\end{equation}
In order to obtain the long range dilaton field we will need
the asymptotic form of a Green's function, satisfying Dirichlet boundary
conditions,  for the equation:
\beq
\label{gdir}
{1 \over 2 \kappa_5^2} \nabla^2 G_D(\vec{x}_{\perp},z) = {z^5 \over R^5}
\delta(z- \tilde{z}) \delta(\vec{x}_{\perp} - \vec{x}_{a\perp}),
\eeq
where the right hand side contains a delta function transverse to the
direction in which the string extends.  This is readily solved to give:

\beq
\label{formdir}
G_D(\vec{x},z)=-{4 \pi \over N^2} {z^4 \tilde{z}^4 \over  
[\tilde{z}^2 + \vert \vec{x}_{\perp} - \vec{x}_{a\perp}\vert^2]^3}.
\label{eq:greens}
\eeq
(This can can also be obtained from our earlier results for the
D-instanton by integrating (\ref{soldilaton}) with respect to the
positions $t_a$ and $x^1_a$ and multiplying by the appropriate
normalization.)  Now it follows from (\ref{dilcoupling}) and
(\ref{formdir}) that
\beq
\label{phifstring}
\phi_F = {g_{YM}\over N^{3/2}} {z^4 \tilde{z}^2 \over
[\tilde{z}^2 + \vert \vec{x}_{\perp} - \vec{x}_{a\perp}\vert^2]^3} \\
\eeq
and
\beq
\label{phidstring}
\phi_D=-\frac{4\pi}{g_{YM}N^{3/2}} \frac{z^4 \tilde{z}^2}
{[\tilde{z}^2 + \vert\vec{x}_{\perp} - \vec{x}_{a\perp}\vert^2]^3}.
\eeq
These asymptotic fields yield the expectation values
\begin{eqnarray}
\langle {\rm Tr}F^2 \rangle_{F} &=& \frac{2g_{YM}^3 N^{1/2}}{\pi^2}
\frac{\tilde{z}^2}{[\tilde{z}^2 + \vert \vec{x}_{\perp}
                           - \vec{x}_{a\perp}\vert^2]^3} 
\label{fflux}\\
\langle {\rm Tr}F^2 \rangle_{D} &=& -\frac{8g_{YM} N^{1/2}}{\pi}
\frac{\tilde{z}^2}{[\tilde{z}^2 + \vert \vec{x}_{\perp}
                                  - \vec{x}_{a\perp}\vert^2]^3}
\label{dflux}
\end{eqnarray}
We see that the boundary configuration corresponding to a static
string in the bulk is spread over a region with scale size
$\tilde{z}$.  This analysis can be generalized to a slowly moving
string by using retarded Green's functions instead of
(\ref{eq:greens}); $\tilde z$ is then replaced by its value at retarded time.

\paragraph{Scale-radius duality and bulk locality: }
We have learned that a bulk string placed at $\tilde z$ is dual to a
flux tube spread over a region with characteristic scale $\tilde z$ --
another example of scale-radius duality.  In the bulk a string will
fall towards the horizon (large $\tilde z$) to minimise gravitational
potential energy. Correspondingly, the gauge field strength in the
boundary theory will spread out to minimize gradient energy,
asymptotically going to zero.  The AdS/CFT correspondence implies
that the equation governing the spreading in the boundary is the
geodesic equation for strings in the bulk.  At present it is difficult
to analyze this directly from the boundary perspective, but we gain
some insights from~\cite{Gava},
\footnote{We thank A. Sen for bringing this reference to our
notice.} where a system of $p+2$ and $p$ branes in flat space is
studied.  The authors found that a D-string can be included in a
D3-brane $SU(N)$ gauge theory as a $Z_N$ flux after compactifying two
directions transverse to the D-string.  Their analysis showed that the
minimum energy configuration with fixed $Z_N$ flux is pure gauge, 
with vanishing field strength.  Taking a large
compactification radius, this agrees with the picture in
(\ref{fflux}-\ref{dflux}) where $\langle Tr(F^2)\rangle$ vanishes as
$\tilde z \rightarrow
\infty$ .  For a classical, static field configuration which is purely
electric or magnetic, $(1/g^2_{YM}) {\rm Tr}F^2$ is proportional to
the energy density.\footnote{Since we do not expect classical
Yang-Mills theory to be accurate in this context, the following
discussion is meant only to indicate the qualitative behavior of the
field configuration.}  One expects that the total energy for a field
configuration of fixed flux and size $\tilde z$ goes like $1/{\tilde
z}^2$, in agreement with the bulk potentials (\ref{potentials}).


Two strings at large radial separations do not interact very much.
This feature can be seen by examining the collective coordinates of
the bulk solitons -- each string has an approximately independent set.
On the boundary the corresponding statement is that flux tubes of very
different scale sizes have approximately independent collective
fluctuations, even when they have coincident centers.\footnote{Of
course, an object like a flux tube, not being a stable soliton, does
not have collective coordinates in the usual sense.  By ``collective
coordinates'' we mean fluctuations of the boundary fields that
preserve the overall scale and shape of the tube.  For example, we can
imagine endowing the tube with a ripple in its shape or a transverse
velocity.  Such motions would occur on time scales distinct from the
rate of spreading of the tube.}  The interactions of bulk strings are
also causal in the classical limit.  For instance, a fluctuation on
one string will only affect the other after a time lag.  This
translates into a typical interval required for the spread of boundary
fluctuations from one scale to another.  As in the case of
D-instantons, the separation of boundary collective fluctuations and
the time lag for interactions between scales are only expected to
emerge in an approximate sense.  At a more fundamental level, the
exact dynamics dictated by the CFT description will imply deviations
from bulk locality and causality.

\section{Dilaton Wave  Packet}
\label{sec:dilapac}

Finally we study massless dilaton wavepackets in the bulk of
$\ads{5}$.  In previous sections we studied pointlike sources and
found that the bulk position translated into a boundary scale.  The
situation is more complicated for wavepackets because the bulk object
already has a characteristic scale which will also get reflected on
the boundary.  The correct approach is to study a family of objects
related in the bulk by the AdS isometry (\ref{eq:isom}).  For the
D-instanton and string probes, this isometry simply translates the
bulk obects.  The component of the translation parallel to the
boundary becomes a translation in the CFT, while the radial
translation becomes a {\em spatial} rescaling.  We will see that the
isometry (\ref{eq:isom}) both translates and changes the size of
dilaton wavepackets.  This is reflected in the boundary theory in the
{\em spatio-temporal} width of $\langle Tr(F^2) \rangle$.


\paragraph{Bulk wavepackets:}  A normalizable dilaton
wavepacket can be constructed by superimposing mode
solutions~\cite{avisetal,brfreed,mez,us}:
\beq
\label{wavepckt}
\phi_{\lambda}(z,\vec x,t) = \int d^3k d\rho ~ C_{\lambda}(\vec k, \rho) ~ 
(\rho  z)^2  J_2(\rho z)  ~ e^{i(\vec k \cdot \vec  x - \omega t)}. 
\eeq
Here $J_2$ is a Bessel function and $\omega^2 - \rho^2 - \vec k^2 =
0$.  The profile $C_\lambda$ is a Gaussian centered at $(\lambda \vec
k_0, \lambda \rho_0)$ with a width $\sigma \lambda^2$:
\beq
\label{sprofile}
C_{\lambda}(\vec k, \rho) = {1 \over \lambda^4}~ a~ e^{[-{(\vec k -
\lambda \vec k_0)^2 +(\rho - \lambda \rho_0)^2 \over 2
\lambda^2  \sigma} -i {\rho \alpha \over \lambda}] },
\eeq
With this definitition, wavepackets with different values of $\lambda$
are related by the radial isometry (\ref{eq:isom}):
\beq
\label{paranew}
\phi_{\lambda}(z,\vec x,t) =\phi_{(\lambda = 1)}(\lambda  z,  \lambda \vec x, 
\lambda t). 
\eeq
So we expect to see a manifestation of the scale-radius duality in the
dual boundary dynamics.

For sufficiently early or late times\footnote{We need $\vert \lambda t
\vert \gg \omega_0/\rho_0^2$ and $\vert \lambda  t \vert \gg \omega_0
/\sigma$.} we can use the stationary phase approximation and the
asymptotic form of the Bessel function for large arguments to study
(\ref{wavepckt}).   We find, self-consistently, that that at large
$|t|$, the packet $\phi_\lambda$ is centered at:
\beq
\label{fcenter}
\vec x = {\vec k_0 \over  \omega_0} t
~~~~~~~;~~~~~~~ z={\rho_0 \over \omega_0}|t|+
{\alpha \over \lambda},
\eeq
As $\lambda$ increases, the center of the wave packet moves radially
away from the horizon.  At large times the detailed form of the
wavepacket is still somewhat complicated but the key features can be
understood by setting $\alpha=0$. This gives:
\beq 
\label{farwpt}
\phi_{\lambda}(z, x_i,t)=  a  {\sqrt \sigma }~
(\rho_0 z)^{3/2}~({\omega_0 \over t})^{3 \over 2} e^{-{\sigma 
\lambda^2\over 2 } (\vec x \cdot {\hat  \zeta_0} - t)^2} ~ e^{-{1 \over 2} {x_T^2
\omega_0^2 \over  t^2 \sigma} }~ e^{{i  \over 2}{x_T^2  \lambda \omega_0
\over  t}}  e^{i(\lambda \vec \zeta_0 \cdot  \vec x -  \lambda \omega_0 t)}. 
\eeq
Here $\vec x$ is the 4-vector $(z, x_i)$, $\vec \zeta_0$ is the
4-vector, $(\rho_0, \vec k_0)$ and $x_T$ stands for the spatial
distance in 4 dimensions transverse to the wave packet's momentum
along $\vec \zeta_0$.  So, as $t \rightarrow -\infty$ the wave packet
is a shock wave that emerges from the horizon travelling along $\hat
\zeta$.  Initially its energy is large and it travels like a massless
particle along the light cone. But with time the pull of gravity gets
stronger and begins to reflect the wavepacket back.  As this happens
the shock wave contracts into a localized lump in the direction
perpendicular to its motion.  At this stage (\ref{farwpt})
is no longer valid.  Eventually the state turns around completely,
gathering itself into a shock wave again, this time hurtling towards
the horizon in the far future and spreading out transverse to its
direction of motion.  This spreading transverse to the direction of
motion will be familiar to reader as the standard behaviour of
relativistic wavepackets in flat space.

In (\ref{farwpt}) we had set $\alpha=0$; reinstating it does not
change the qualitative features of the wave packet.  In particular the
widths $\sigma
\lambda^2$ and ${\omega_0^2 \over t^2 \sigma}$ , which govern the 
spreading parallel and perpendicular to the direction of motion stay
the same.  Thus, shock waves closer to the horizon (smaller $\lambda$
from (\ref{fcenter})) are also more spread out along their direction
of motion.

\paragraph{Boundary Description:} Using the results of
Sec.~\ref{sec:relbb} we relate the asymptotic behaviour of the dilaton
wave packet to the expectation value of $\langle Tr F^2 \rangle$:
\beq
\label{trf2} 
F^2(\vec{x}) \equiv {1 \over 4 g_{YM}^2} <\lambda|Tr(F^2(\vec
x))|\lambda> = \int d^3k \, d\rho ~ 4 C_{\lambda}(\vec k, \rho) ~ \rho^4~
e^{i(\vec k \cdot \vec x - \omega t)},
\eeq
Here the states $|\lambda \rangle$ are related to each other by a scale
transformation dual to the isometry~(\ref{eq:isom}).
Two limiting cases are instructive: $\rho_0 \gg \vert \vec k_0 \vert$
and $\rho_0 \ll \vert \vec k_0 \vert$.

When $\rho_0 \gg \vert \vec k_0 \vert$ the wavepacket starts at early
times with most of its momentum in the radial direction.  Then
(\ref{trf2}) gives:
\begin{eqnarray}
\vert \lambda t\vert \gg {\omega_0 \over
\sigma}:~~~~~~
F^2(\vec{x}) &\propto&
\left({\omega_0 \over t}\right)^{3  \over 2} \: e^{-{\sigma \lambda^2
  \over 2} 
(t + {\alpha \over \lambda})^2} 
\: e^{-{1 \over 2} {\vec x^2 \omega_0^2 \over t^2
\sigma }}
\:  e^{{i  \over 2}{\vec  x^2 \omega_0 \lambda \over  t}}
\:  e^{-i \omega_0  (\lambda t + \alpha)}
\label{paraf212} \\
\vert \lambda t\vert \ll {\omega_0
\over \sigma}:~~~~~~
F^2(\vec{x}) &\propto&
e^{-{\sigma \lambda^2  \over 2}
(t + {\alpha \over \lambda})^2} ~e^{-{1 \over 2}({\vec  x^2  \sigma
\lambda^2})} \:
e^{-i \omega_0  (\lambda t + \alpha)}
\label{stf1}
\end{eqnarray}
At early and late times ($\vert \lambda t\vert \gg {\omega_0 \over
\sigma}$) the bulk state looks like a shock wave moving in the radial
direction and at small intermediate times ($\vert \lambda t\vert \ll
{\omega_0 \over \sigma}$) the bulk state is being reflected by the AdS
geometry and turned around.

When $\rho_0 \ll \vert \vec k_0 \vert$ the wavepacket starts at early
times with most of its momentum parallel to the boundary.  Then we
find:
\begin{eqnarray}
\vert \lambda t\vert \gg {\omega_0 \over
\sigma}:~~~~~~
F^2(\vec{x}) &\propto&
({\omega_0 \over t})^{3 \over 2} \: e^{-{\sigma \lambda^2  \over 2}
(\vec x \cdot \hat k_0 - t)^2} 
\: e^{-{1 \over 2} {x_T^2 \omega_0^2 \over
t^2 \sigma}}  
\: e^{{i  \over 2}{ x_T^2 \omega_0 \lambda \over  t}}
\: e^{i \lambda (\vec k_0 \cdot\vec x - \omega_0 t)}
\label{paralastf2}
\\
\vert \lambda t\vert \ll {\omega_0
\over \sigma}:~~~~~~
F^2(\vec{x}) &\propto&
 e^{-{\sigma \lambda^2  \over 2} 
(\vec x \cdot \hat k_0 - t)^2} 
\: e^{-{1 \over 2} (x_T^2 \lambda^2
\sigma)} 
\: e^{i \lambda (\vec k_0 \cdot\vec x - \omega_0 t)},
\label{stf2}
\end{eqnarray}
Here $x^2_T = \vec{x}^2 - (\vec{x} \cdot \hat{k}_0)^2 +
({\alpha\over\lambda})^2$.  At early and late times ($\vert \lambda
t\vert \gg {\omega_0 \over \sigma}$) the bulk state looks like a shock
wave moving parallel to the boundary.

\paragraph{Scale-radius duality: }
We have just derived the boundary description of a class of
wavepackets $\phi_\lambda$ that are related by the radial isometry
(\ref{eq:isom}).  We learned from (\ref{fcenter}) that the
characteristic spatial center of the packet depends linearly on
$1/\lambda$.  Packets that are closer to the horizon also have
bigger bulk widths.  The basic lesson we learn from (\ref{paraf212}) -
(\ref{stf2}) is that these characteristic bulk features map on the
boundary to a characteristic {\em spatio-temporal} scale.  Packets
which are characteristically closer to the horizon (and spatially
wider) map to boundary fields with a greater spread in space and
time.

To see this, first consider the case $\rho_0 \gg |\vec{k}_0|$.  The
bulk packet starts as a radial shockwave coming from the past horizon,
reflects at intermediate times and returns as a shockwave to the
future horizon.  From (\ref{paraf212}) the boundary $F^2$ starts with
a very small amplitude and a huge spread in the spatial directions.
At intermediate times (\ref{stf1}) tells us that $F^2$ is a Gaussian
in space and at late times the amplitude decreases again while $F^2$
spreads in space.  The temporal profile, like the intermediate time
spatial profile, is Gaussian with a scale set by $\lambda$.  Since
$\lambda$ also indexes the radial isometries relating bulk packets, we
are once again seeing a scale-radius duality.

As another example, consider the case $\rho_0 \ll |\vec{k}_0|$.  Now
the bulk packet starts as a shockwave parallel to the boundary, is
reflected at intermediate times and returns as a shockwave to the
future horizon.   From (\ref{paralastf2}) and (\ref{stf2}) we learn
that the boundary $F^2$ is a shockwave spreading out spatially
in the directions perpendicular to the motion at early and late times.
Other than this, the behaviour is exactly parallel to the case $\rho_0
\gg |\vec{k}_0|$.  The profile in time, like the intermediate time
spatial profile, is a Gaussian with a scale set by $\lambda$.


The analysis of this section is only a first step in a more complete
study.   For example, it would be interesting to understand how the
bulk scattering of two shock waves in mirrored in the boundary
theory. Understanding this would help uncover how bulk locality
emerges from the boundary description.

\section{Discussion and conclusions}

\subsection{Bulk versus boundary dynamics}
\label{sec:exclusion}

\paragraph{The classical limit:}
Much of this article has addressed the CFT description of classical
bulk probes.  These states are ``classical'' because they contain a
very large number of particles, and propagate in backgrounds with
small curvature.  There is a subtlety in defining such a limit in view
of the ``stringy exclusion principle'' advocated in~\cite{jaex} for
$\ads{3}$.  Given the AdS/CFT correspondence, this ``exclusion
principle'' imposes a bound on the occupancy of certain bulk states.
Nevertheless, as $N$ increases, the maximum occupancy increases also.
So, in the large $N$ limit our considerations are valid.

\paragraph{Hairy holography: }
We have provided an explicit prescription for relating states in the
bulk and boundary theories.  Roughly speaking, the prescription works
because normalizable modes in the bulk extend to the boundary and
serve as a kind of ``hair'', determining the boundary state.  Each
mode falls off exponentially (in physical distance), but the volume of
the boundary grows exponentially as well allowing for a significant
effect.  We have found that the asymptotic value of the ``hair''
appears directly in the expectation values of operators in the
boundary state, giving a precise realization of the holographic
proposal of `t Hooft and Susskind~\cite{holog}. Our analysis has been
mainly at a linearised level and back reaction on the metric was
neglected.  We hope to extend our work in the future to situations
like the formation of a black hole where the nonlinearities of gravity
are more important.

\paragraph{S-matrix, bulk commutators and local physics: }
Using the operator relation between bulk and boundary fields, we can
reiterate some points made in~\cite{banksgreen,us}.  It is apparent
that transition amplitudes between physical states in the bulk can be
computed from the boundary theory.  Specifically, prepare a bulk state
$|\Psi \rangle = a_{k_1}^\dagger \cdots a_{k_n}^\dagger |0\rangle$ and
evolve it forward in time as $|\Psi
\rangle \rightarrow e^{-iHt}|\Psi \rangle$.  Precisely the same
operation can be performed in the boundary theory: the expansion
(\ref{Oop}) and the equivalence $a_k=b_k$ allow one to prepare the
intitial state, and time evolution with respect to the CFT Hamiltonian
is identified with time evolution in the bulk.  Thus transition
amplitudes in the boundary theory can be reinterpreted as 
amplitudes in the bulk.

An unusual feature of this map is that the radial coordinate in the
bulk does not appear in the boundary mode expansion (\ref{Oop}).  This
makes it difficult to check aspects of bulk locality such as the
commutation of operators at spacelike separation.  However, as we have
explicitly shown, the boundary theory has access to data on the radial
position of localized bulk probes in the characteristic scale of their
boundary images.  As we have discussed, classical locality of the bulk
physics is mapped, at large $N$, into the approximate independence of
collective fluctuations of objects at different boundary scales.  So
the breakdown of locality in quantum gravity should be understood in
terms of incomplete decoupling of scales in the boundary theory at
finite $N$.  We are investigating this issue and hope to report on it
elsewhere.

\subsection{Holographic description of horizons}
We have accumulated enough tools to suggest how spacetime causal
structure will be reflected in the boundary theory.  Here we 
present a qualitative discussion - further details will appear
elsewhere.

\paragraph{Black holes and thermal states: } AdS-Schwarzchild
black holes~\cite{hawkingpage} and the BTZ black hole~\cite{BTZ} (see
\cite{carlip} for a review) both have
maximally extended solutions with two asymptotic regions, each with a
timelike boundary at spatial infinity.\footnote{The disconnected
topology of the boundary in the BTZ case can be seen directly from its
orbifold construction~\cite{garydon}.}  In such spacetimes, the bulk
Hilbert space is a product of two identical copies, each accessed by a
single asymptotic region~\cite{birrelldavies}.  For example, thermal
states such as the Hartle-Hawking vacuum are written as correlated
tensor products of states:
\be
	|{\rm HH}\rangle = 
	\sum_{n} e^{-\beta \omega_{n}} 
	|n,\omega_n\rangle_I \otimes
	|n,\omega_n\rangle_{II}\ ,
\ee
where Hilbert spaces $I$ and $II$ are formally identical.
Tracing over one copy in the product produces the thermal ensemble
accessible to the other asymptotic observer.  In fact, this
construction is a standard method of describing real-time,
finite-temperature field theories; one studies operators that access
one Hilbert space that is correlated appropriately with another.  The
auxiliary Hilbert space then functions as an external bath which
thermalizes the system.  The situation is entirely parallel from the
boundary perspective.  The boundary of spacetime has two disconnected
components and so the CFT Hilbert space factorizes into a product of
two identical pieces.  The choice of bulk vacuum is reflect in the CFT
vacuum as discussed in Sec.~\ref{sec:operator} and
Appendix~\ref{app:corrs}.  Tracing over one boundary component leaves
a thermal state accessible to one asymptotic region.

\paragraph{Scale-radius duality: }
We would like to use the scale-radius duality discussed in this paper
to study motions towards the horizon from the boundary perspective.
For pure AdS, scale-radius duality originates in the dual action of
the bulk radial isometry and the boundary scale transformations.  In
fact, these are not symmetries of the black hole. The curvature of
Schwarzchild and the discrete identifications of BTZ break the
isometry group, and the corresponding thermal boundary state breaks
conformal invariance.  Nevertheless, there are arguments that the
bulk-boundary duality continues to hold since the spacetime is {\em
asymptotically} AdS~\cite{juanads,holowit,itzhakietal,thermwitt}.
What is more, BTZ black holes {\em locally} enjoy the same symmetries
as AdS and so motions of local probes continue to map to motions in
boundary scale.  (These points can be made quantitatively using the
methods of this paper, as we hope to discuss elsewhere.)

\paragraph{Horizons from the boundary perspective: }
Armed with the scale-radius duality, we introduce a bulk probe that
starts near the boundary and falls towards the horizon.  From the
boundary perspective, the position of the horizon is represented by
the thermal scale.  The boundary probe starts life at a very high
scale much above the thermal bath.  This separation of scales allows
it to spread unimpeded as though the bath was absent.  As the bulk
object falls through the horizon, its boundary dual reaches the
thermal scale.  Falling through the horizon in the bulk is reflected
on the boundary as thermalization due to interaction with the thermal
bath.  Note that this does not mean that the probe state has ``mixed''
with the density matrix describing the black hole.  Rather,
interactions with the thermal bath make the probe state look like a
typical state in the ensemble.  As the bulk object penetrates to the
singularity increasing the black hole mass, thermalization of its
boundary dual raises the boundary temperature.  The horizon as a
causal construct preventing extraction of information is only ``real''
to the degree that thermalization obscures the history of a state.

\paragraph{Black holes from collapse: }
We can also consider black holes formed from collapsing shells of
matter.  Again, from the boundary perspective, the state will spread
out until it reaches the scale characteristic of the temperature of
the black hole it has created.  The degree to which the resulting
horizon is sharp will be the degree to which the final state is
difficult to measure due to the complicated way that the information
about the configuration is spread out over modes at low spatiotemporal
scales.  The causal structure of the black hole appears as a
statistical phenomenon and is precise in the thermodynamic limit.

\subsection{Conclusions}
In summary, we have developed techniques for describing bulk probes
from the boundary perspective.  Using our methods, properties of
solutions to the bulk equations of motion can be translated into
properties of states and expectation values in the boundary theory.
We have argued that there is a map between the quantized mode
expansions of bulk fields and boundary operators and used our approach
to demonstrate a scale-radius duality for several probes.  Finally, we
outlined the application of our methods to the study of black hole
causal structure.  This work constitutes a preliminary attempt to
address the emergence and eventual breakdown of local spacetime from
the gauge theory perspective towards quantum gravity.

\vspace{0.2in} {\bf Acknowledgments:} 
V.B. is supported by the Harvard Society of Fellows and by NSF grant
NSF-PHY-9802709.  P.K. is supported in part by DOE grant
DE-FG03-92-ER40701 and by the DuBridge foundation.  A.L. is supported
by NSF grant NSF-PHY-9802709. S.T. is supported in part by DOE grant
DE-AC02-76CH0300.  V.B. is grateful to the organizers of the Amsterdam
Summer Workshop on String Theory and Black Holes and to the Xerox Palo
Alto Research Center for hospitality while this work was being done.
We are grateful to Ofer Aharony, Rajesh Gopakumar, Esko Keski-Vakkuri,
Samir Mathur, Nikita Nekrasov, John Preskill, John Schwarz, Ashoke
Sen, Andy Strominger and Barton Zweibach for interesting discussions.
While this work was in progress we learned of~\cite{tometal} where
related issues are discussed.

\appendix

\section{Correlators and propagator ambiguities}
\label{app:corrs}
In this appendix we will give a more careful discussion of the methods
illustrated in Sec.~\ref{sec:relbb}.  In particular we will discuss 
the meaning of the ambiguities in defining the bulk-boundary
propagator for Lorentzian spacetimes.  We will continue to work in
Poincar\'e coordinates (\ref{eq:poinmet}).

\subsection{Euclidean signature}
As in Sec.~\ref{sec:relbb} we study the bulk-boundary correspondence
(\ref{eq:bbrel}) for a classical massive scalar with action
(\ref{action}) that couples to an operator with dimension $2h_+$
(\ref{eq:hpm}).   In Sec.~\ref{sec:relbb} we worked in position space
and followed the procedure of~\cite{holowit}  relating the boundary
contribution to the bulk action to CFT correlators.  In fact, as
discussed in~\cite{freedcorr} this procedure violates Ward identities
and must be modified.  The correct algorithm is to evaluate a suitably
normalized bulk action at $z=\epsilon$ prior to taking $\epsilon
\rightarrow 0$.

\paragraph{Improved procedure in momentum space} 
The improved procedure of~\cite{freedcorr} is easiest to implement in
momentum space.   The unique solution to the wave equation $(\Box
-m^2)\phi = 0$ with momentum $\vec{k}$ parallel to the boundary is the
Bessel function~\cite{gkp,freedcorr,us}:
\begin{equation}
	\phi \propto e^{i\vec{k}\cdot\vec{x}} z^{\frac{d}{2}} K_\nu (|k| z)
		\phi_0 (\vec{k})\ .
\end{equation}
At the boundary $z \rightarrow 0$, $\phi \rightarrow A(z) + B(z)$
where $A = z^{2h_-} (1 + \cdots)$ and $B = z^{2h_+} (1 + \cdots)$ and
the ellipses indicate series in even powers of $z$.\footnote{For
integral $\nu$ there are some log terms also, but they do not change
the basic argument.}  For $m^2>0$ this is divergent as $z \rightarrow
0$ and requires regulation.  According to~\cite{freedcorr} we work at
$z = \epsilon$ with the normalization $\phi = C(\epsilon)
\phi_0(\vec{k}) e^{i\vec{k}\cdot\vec{x}}$.  The authors
of~\cite{freedcorr} set $C(\epsilon) = 1$, but since the scaling of
$\phi$ as it approaches the boundary is important, we choose
$C=\epsilon^{2h_-}$:
\begin{equation}
	\phi(z,\vec{k}) = \epsilon^{2h_-} \frac{ z^{\frac{d}{2}} K_\nu (|k| z)}
		{\epsilon^{\frac{d}{2}} K_\nu (|k|\epsilon)} 
         \phi_0(\vec{k})e^{i \vec{k} \cdot \vec{x}}
\label{eq:momsource}
\end{equation}
The bulk action (\ref{action}) then reduces to a boundary
term~\cite{gkp,freedcorr}:
\begin{equation}
	S = \lim_{z\to \epsilon} z^{1-d} \delta(\vec{k} + \vec{k}')
	\phi_0(\vec{k}') \epsilon^{2h_-} 
		\frac{(|k'|z)^{\frac{d}{2}} K_\nu (|k'|z)}
			{(|k'|\epsilon)^{\frac{d}{2}} K_\nu (|k'|\epsilon)}
	\partial_z \epsilon^{2h_-} 
		\frac{(|k|z)^{\frac{d}{2}} K_\nu (|k'|z)}
			{(|k|\epsilon)^{\frac{d}{2}}K_\nu (|k'|\epsilon)}\ .
\label{eq:momcorr1}
\end{equation}
As we discussed, $z^{d/2} K_\nu(z) \rightarrow A(z) + B(z) = z^{2h_-}
(1 + \cdots) + z^{2h_+} (1 + \cdots)$.  Putting this in
(\ref{eq:momcorr1}) as $\epsilon \rightarrow 0$ yields some singular
contact terms that we drop and a finite term arising from the mixture
of the $A$ and $B$ terms in the numerator and denominator.  These give
the expected behaviour of the 2-point function:
\begin{equation}
	<\CO (\vec{k}') \CO(\vec{k})> \propto \delta (\vec{k}' + \vec{k})
		|k|^{2\nu}\ .
\label{eq:momtwo}
\end{equation}
The position space procedure used in Sec.~\ref{sec:relbb} gives
different normalizations because the bulk-boundary propagator used
there approaches a delta function at the $z=0$ boundary rather than at
$z=\epsilon$.

\paragraph{Operator expectation values}
From Sec.~\ref{sec:relbb} we know that turning on source in the
boundary theory will lead to nontrivial operator expectation values.
These are given by the first variation of $S(\phi)$:
\begin{equation}
	\delta S(\phi) = \int_{z\to 0} d\Sigma^\mu \p_\mu\phi
		\delta\phi\ .
\label{eq:firstvar}
\end{equation}
where, as in Sec.~\ref{sec:relbb} we have added a small perturbation
of the form $\delta\phi = z^{2h_-}\delta\phi_0$.
In momentum space, with the cutoff procedure
prescribed above, 
$\delta\phi(z=\epsilon,\vec{k}) = \epsilon^{2h_-} \delta\phi_0(\vec{k})$.
Combining this with $\p_z \phi$ at $z=\epsilon$,
using (\ref{eq:firstvar}) and dropping contact terms as before,
we get a finite one-point function:
\begin{equation}
	\langle \CO(\vec{k})\rangle_{\phi_0(\vec{k})}
	= \langle\CO(\vec{k})\CO(-\vec{k}')\rangle
		\phi_0 (\vec{k}')
\end{equation}
where the two-point function was given in (\ref{eq:momtwo}) and is
evaluated here in the absence of a source.
So the interesting part
of the one-point function -- the part which does
not come from coincidence of the source and
the operator insertion -- arises
from the subleading part of the source term
which scales as $z^{2h_+}$ at the boundary.
The bulk of this paper rests on the independent
specification of this subleading part near $z=0$ via the addition 
of bulk probes and the consequent modifications of operator
expectations on the boundary.

\subsection{Lorentzian signature}
We again begin by discussing free, massive scalar fields.  In
Lorentzian signature, normalizable solutions to the wave equation
exist; so specifying the fields at $z=0$ does not uniquely specify the
field configuration in the bulk.  As discussed in~\cite{us} the
normalizable solutions form the bulk Hilbert space which is dual to
the space of boundary states.  There is also a spectrum of
non-normalizable modes that implement boundary conditions and are dual
to boundary sources.

In the supergravity effective action, the normalizable solutions can
appear in two places.  First, classical field theory in the bulk
involves expanding the bulk action around classical, normalizable
backgrounds.  We will see that this corresponds to turning on
expectation values for CFT operators.  Secondly, the bulk-boundary
propagator is not unquiely specified by the asymptotic behaviour
$z^{2h_-}\delta^{d}(\vec{x}-\vec{x}')$ as $z\to 0$, since a
normalizable solution vanishing at the boundary can always be added to
it.   This ambiguity is related to the choice of vacuum for the theory
and will affect the correlation functions.

Working in momentum space and in Poincar\'e coordinates
(\ref{eq:poinmet}), we write $\vec{k}=(\omega,\vec{q})$ for the
momentum parallel to boundary ($\vec{y} = (t,\vec{x})$).  For $k^2 > 0$
(spacelike momenta) the solutions are identical to the Euclidean case
and there is no normalizable solutions.  For $k^2 < 0$, there are two
solutions which are smooth in the interior~\cite{us}.  One solution is
\be
	\phi^{-}(\vec{k},z) \propto z^{\frac{d}{2}} J_{-\nu} (|k|z)
e^{i\vec{k}\cdot \vec{y}} 
\ee
when $\nu$ is nonintegral.\footnote{When $\nu$ is integral,
$\phi^{-}(\vec{k},z) \propto z^{\frac{d}{2}} Y_\nu (|k|z)
e^{i\vec{k}\cdot
\vec{y}}$} This solution is not normalizable (for $\nu > 1$ --
see~\cite{us} for a discussion of the case $\nu < 1$)); it behaves as
$z^{2h_-}$ when $z\to 0$.  So the mode is non-fluctuating and is dual
to a source term at the boundary.  An independent solution is:
\be
	\phi^+(\vec{k},z) = z^{\frac{d}{2}} J_\nu (|k|z) e^{i\vec{k}\cdot
\vec{y}}\ ,
\ee
which is normalizable and behaves as $z^{2h_+}$ when $z\to 0$.

\paragraph{Operator expectation values: }
The most general solution $\phi(\vec{k},z)$ which is asymptotic to to
$\epsilon^{2h_-}\phi_0(\vec{k}) e^{i\vec{k}\cdot\vec{y}}$ when
$z=\epsilon$ is:
\be
	\phi(\vec{k},z) = \Phi(\vec{k}) e^{i\vec{k}\cdot
\vec{y}}\phi^+ (\vec{k},z) + \epsilon^{2h_-}
		\frac{ \phi^- (\vec{k},z) + A(\vec{k}) \phi^+(\vec{k},z)}
		{\phi^-(\vec{k},\epsilon) + A(\vec{k}) \phi^+(\vec{k},\epsilon)}
	e^{i\vec{k}\cdot \vec{y}}\phi_0(\vec{k})\ .
\ee
Here $\Phi(\vec{k})$ is the Fourier component of
the classical field configuration we wish to study and 
\be
	K_A(\vec{k},\epsilon,z) = \epsilon^{2h_-} \frac{ \phi^-
		(\vec{k},z) + A(\vec{k}) \phi^+ (\vec{k},z)} {\phi^-
		(\vec{k},\epsilon) + A(\vec{k}) \phi^+
		(\vec{k},\epsilon)}
\label{eq:lorbbgf}
\ee
is the bulk-boundary Green's function.  As we can see, $K_A$
has an ambiguity which is encoded by $A(\vec{k})$.
Once again, we insert this into the (quadratic)
action and integrate by parts.\footnote{In fact the integration by 
parts gives an oscillating surface term at $z \rightarrow \infty$.
This should be understood in terms of an infrared cutoff in the gauge theory.}
The result is:
\begin{eqnarray}
	S_A(\Phi,\phi_0) = \frac{1}{2}\int_{z=\epsilon} \, d^dk \, d^dk'
	 \, 	\delta^d(\vec{k}+\vec{k}') \, z^{1-d} \, \partial_z \left[ 
	\Phi(\vec{k})\, \phi^+(\vec{k},z)\,   K_A(\vec{k}',\epsilon,z) \, \phi_0(\vec{k}') \right.
\nonumber \\
 \left.	+ \phi_0(\vec{k}) \, K_A(\vec{k}',\epsilon,z) \, \phi_0 (\vec{k}')\right]
\end{eqnarray}
From this we can read off the one-point function:
\be
	\langle \CO(\vec{k}) \rangle_{\phi_0} = \lim_{z\to\epsilon}
	\partial_z \left[ 
       \left(\Phi(-\vec{k}) \phi^+(-\vec{k},z) + \phi_0(-\vec{k})\right) 
	K(\vec{k},\epsilon,z) \right]
\label{eq:lorone}
\ee
When $\phi_0 = 0$, it is easy to see that
\be
	\langle \CO(\vec{k}) \rangle = {d \over 2} \Phi(-\vec{k}) \,
\ee
The two-point function is:
\be
	\langle \CO(\vec{k})\CO(\vec{k}') \rangle_{\phi_0}
		= \langle \CO(\vec{k}) \rangle_{\phi_0}
		\langle \CO(\vec{k}') \rangle_{\phi_0} + 
		\lim_{z\to\epsilon}
		\delta^d(\vec{k}+\vec{k}') \p_z K(\vec{k},\epsilon,z)
\label{eq:lortwo}
\ee
In this case, the parts of $K$ which behave as $z^{2h_+}$ and
$z^{2h_-}$ are explicitly labeled by $\phi^+$ and $\phi^-$.  As in the
Euclidean case, after dropping contact terms the finite piece as
$\epsilon\to 0$ comes combinations of $\phi^-$ terms in the numerator
and $\phi^+$ terms in the denominator (and vice-versa)
of~(\ref{eq:lorbbgf}).  Thus the connected part of the two-point
function will depend quite strongly on $A(\vec{k})$.

\paragraph{Interpretation of propagator ambiguities: }
The dependence of these correlators on $\Phi$ is easy to understand.
Turning on such a classical background implies that the bulk theory is
in a ``coherent'' state in which many modes have an expectation value.
The map between bulk and boundary states discussed in
Sec.~\ref{sec:relbb} implies that the CFT should also be in a ``coherent''
state built from the modes of the dual boundary operator.   Not
surprisingly, the CFT operator has an expectation value in this state.
Roughly speaking, the expectation value is dual the part of $\phi$
that behaves as $z^{2h_+}$ at the boundary.

In addition to the freedom to specify a classical background/coherent
state via $\Phi$, there is an ambiguity arising from the freedom to
specify $A(\vec{k})$.\footnote{We are grateful to Tom Banks and Emil
Martinec for a discussion of this issue.}  The origin of the
ambiguity is the freedom to specify a vacuum state.  In standard field
theory, Wick rotation from a Euclidean signature specifies a certain
propagator.  But if we were simply to ask that the two-point function
satisfy the appropriate inhomogenous differential equation, we would
have a much wider range of choices; we could pick some combination of
the advanced, retarded, and Feynman propagators, and we could add
propagators that solved the associated homogenous differential
equation.  Our choice would depend on boundary conditions and the
choice of vacuum.  Here we choose $A$ via analytic continuation from
the Euclidean case; the combination of $J_\nu(|\vec{k}|z)$ and 
$J_{-\nu}(|\vec{k}|z)$
will be that equal to $K_\nu (i|\vec{k}|z)$:
\be
	K_\nu (i|\vec{k}|z) = \frac{\pi}{2}\frac{e^{i\pi\nu/2} 
J_{-\nu}(-|\vec{k}|z) -
		e^{-i\pi\nu/2} J_\nu (-|\vec{k}|z)}{\sin(\nu\pi)}\ .
\ee
In some spacetimes (e.g. those containing black holes) there are
inequivalent vacua defined with respect to different times.
Presumably this would be reflected in the the bulk-boundary
propagator.

\section{The Bulk Response: Another method}
The relation between the dilaton field and the vev of $Tr F^2$ was
obtained in Sec.~\ref{sec:dinst} by considering the response of the
bulk and boundary theories to small perturbations.  An alternate
method for computing the bulk response in the presence of branes
is to do things in the opposite order.  Instead of starting with the
bulk soliton and evaluating the effect of a perturbation, we could 
consider a perturbation of AdS and ask about the change in the action
on including the soliton.  Here we study the D-instaton using the
latter technique.

Consider a perturbation of the dilaton in $\ads{5}$ which reduces to a
delta function at $\vec x = \vec x_a$ on the boundary.  The bulk value
of this perturbation is:
\beq
\label{bpert}
\delta \phi = {6 \over \pi^2} {z^4  \over [z^2 + 
\vert \vec x - \vec x_a \vert^2 ]^4}.
\eeq
The D-instanton is a source for the dilaton and axion and so the
linearized effects of a D-instanton at $z=\tilde{z}$ and $\vec x =
\vec{x}_1$ can be incorporated by adding a term in the action of the
form:
\beq
\label{sdbrane}
S_{Disnt}= 2 \pi \int J  (e^{-\phi} + i \chi)
\eeq
where $J= \delta(x_0-\tilde x_0) \delta (\vec x - \vec x_a)$ stands
for the D instanton source. The coefficient in front of the integral
in eq. (\ref{sdbrane}) is obtained by requiring that the D-instanton
action is given by $S= 2 \pi / g_s$.  Writing $e^{-\phi}= {1\over g_s}
(1 - \delta \phi)$ and susbtituting for $\delta \phi$ from
(\ref{bpert}) gives:
\beq
\label{sdbranestwo}
S_{Disnt}= -{48   \over  4 \pi g_s} {x_0^4  \over [x_0^2 + 
\vert \vec x - \vec x_a \vert^2 ]^4}. 
\eeq
This is exactly equal to the bulk response in (\ref{finaldels}).

This method gives the same bulk response as in Sec.~\ref{sec:dinst}
because we are working in the linearized limit.  An analogy with
electrodynamics is useful. Introduce a small perurbation of the
electrostatic potential surrounding a system of charges.  The energy
of the resulting system can be computed from the action for the
electromagnetic field and the source coupling to the field.  Then the
bulk contribution vanishes by the equations of motion and we are left
with a surface integral like (\ref{dels}).  This method is followed
in most of this paper.  Alternatively, the energy is $E=
\sum_i q_i V_i$ , where $q_i$ and $V_i$ are the charge and the potential
at the position of each charge respectively.  So the change in energy
is determined by the perturbation of the potential at the location of
each charge. This is exactly analogous to (\ref{sdbrane}).

The same reasoning works for a general linear system.  The
supergravity theory under consideration here is certainly not linear,
but the D-branes act as small sources.  Since they carry a charge of
order $ \sim 1/g_s$ the corresponding changes in the supergravity
fields are of order $ {\cal O} (g_s^2 \times 1/g_s)$.  In the large N
limit that we are working in, these changes are of order ${\cal O }
(1/N)$ and are therefore small \footnote{For example, from
eq. (\ref{soldilaton}), eq.(\ref{normc}), we see that the dilaton is
changed from it's asymptotic value by an amount $\delta \phi = {1\over
N} ~{24 \pi\over g_s N}~ { x_0^4 ~\tilde x_0^4
\over [\tilde x_0^2 + \vert \vec x- \vec x_a \vert^2]^4} $. 
 In the large N limit where $g_s N$ is kept fixed this is of ${\cal O}
(1/N)$ .}.  Thus to leading order in $1/N$ (and for purposes of
calculating the one-point functions) one can work with a supergravity
action expanded to quadratic order about the AdS background. The
resulting system is in effect linear and the two methods for
evaluating the bulk response must then agree.



\begin{thebibliography}{10}

\bibitem{juanads}
J. Maldacena,
\newblock ``The large N limit of superconformal field
theories and supergravity'',
\newblock Harvard preprint HUTP-98-A097, hep-th/9712200.


\bibitem{gkp}
S.S. Gubser, I.R. Klebanov and A.M. Polyakov,
\newblock ``Gauge theory correlators from noncritical string theory'',
\newblock Princeton preprint PUPT-1767, hep-th/9802109.

\bibitem{holowit}
E. Witten. 
\newblock ``Anti-de Sitter space and holography'',
\newblock IAS preprint IASSNS-HEP-98-15, hep-th/9802150.

\bibitem{us}
V. Balasubramanian, P. Kraus and A. Lawrence,
\newblock ``Bulk vs. boundary dynamics in anti-de Sitter spacetime'',
\newblock Harvard preprint HUTP-98/A028, Caltech preprint CALT68-2171,
hep-th/9805171



\bibitem{banksgreen}
T. Banks and M.B. Green,
\newblock ``Non-perturbative effects in ${\rm AdS}_5 \times S^5$
string theory and $d=4$ SUSY Yang-Mills'',
\newblock hep-th/9804170.


\bibitem{inst1}
C-S. Chu, P-M. Ho and Y-Y. Wu,
\newblock ``D-instanton in AdS(5) and instanton in SYM(4)'',
\newblock hep-th/9806103.


\bibitem{kogan}
I.I. Kogan and G. Luz\'on,
\newblock ``D-instantons on the boundary'',
\newblock hep-th/9806197.


\bibitem{greenetal}
M. Bianchi, M.B. Green, S. Kovacs and G. Rossi,
\newblock
``Instantons in supersymmetric Yang-Mills and D-instantons in IIB
superstring theory'',
\newblock
hep-th/9807033.

\bibitem{freedcorr}
D.Z. Freedman, S.D. Mathur, A. Matusis and L. Rastelli,
\newblock ``Correlation functions in the ${\rm CFT}_d/{\rm AdS}_{d+1}$
correspondence'',
\newblock MIT preprint MIT-CTP-2727, hep-th/9804058.

\bibitem{avisetal}
S.J. Avis, C.J. Isham and D. Storey,
\newblock ``Quantum field theory in anti-de Sitter space-time'',
\newblock \pr {\bf D18} (1978) 3565.

\bibitem{brfreed}
P. Breitenlohner and D.Z. Freedman,
\newblock ``Positive energy in anti-de Sitter backgrounds
and gauged extended supergravity'',
\newblock \pl {\bf B115} (1982) 197;
\newblock ``Stability in gauged extended supergravity'',
\newblock \ap {\bf 144} (1982) 249.

\bibitem{mez}
L. Mezinescu and P.K. Townsend,
\newblock ``Stability at a local maximum in
higher-dimensional anti-de Sitter space and
applications to supergravity'',
\newblock \ap\ {\bf 160} (1985) 406. 


\bibitem{tometal}
T. Banks, M.R. Douglas, G. Horowitz and E.J. Martinec,
\newblock Talk delivered by T. Banks at Strings '98 and work in progress,
\newblock http://www.itp.ucsb.edu/



\bibitem{klebetal} I.R. Klebanov,
\newblock ``World volume approach to absorption by nondilatonic branes'',
\newblock \np {\bf B496} (1997) 231,
\newblock hep-th/9702076; ~~~~~~~
\newblock S.S. Gubser, I.R. Klebanov and A.A. Tseytlin,
\newblock String theory and classical absorption by three-branes,
\newblock \np {\bf B499} (1997) 217,
\newblock hep-th/9703040.



\bibitem{arvind}
A. Rajaraman,
\newblock ``Two-form fields and the gauge theory description
of black holes'',
\newblock hep-th/9803082.


\bibitem{dastriv}
S.R. Das and S.P. Trivedi,
\newblock
``Three-brane action and the correspondence between N=4 Yang-Mills
theory and anti-de Sitter space'',
\newblock hep-th/9804149.

\bibitem{ferrara}
S. Ferrara, M.A. Lledo and A. Zaffaroni, 
\newblock
"Born-Infeld Corrections to D3 brane
action in $AdS_5 \times S_5$ and $N=4, d=4$ primary superfields'',
\newblock hep-th/9805082.


\bibitem{tseytlin}
A.A. Tseytlin,
\newblock 
``On nonabelian generalization of Born-Infeld action in string
theory'',
\newblock \np {\bf B501} (1997) 41,
\newblock hep-th/9701125.


\bibitem{romansetal}
H.J. Kim, L.J. Romans and P. van Nieuwenhuizen,
\newblock ``Mass spectrum of chiral ten-dimensional
$N=2$ supergravity on $S^5$'',
\newblock \pr\ {|\bf D}32 (1985) 389.



\bibitem{oferwit}
O. Aharony and E. Witten,
\newblock
``Anti-de Sitter space and the center of the gauge group'',
\newblock hep-th/9807205.



\bibitem{Gava} 
E. Gava, K.S.Narain and M.H. Sarmadi,
\newblock "On the bound states of $p-$ and $(p+2)$- Branes'',
\newblock hep-th/9704006.



\bibitem{holog}
G. 't Hooft,
\newblock ``Dimensional reduction in quantum gravity'',
\newblock Salamfest 1993:0284-296, gr-qc/9310026;~~~~~
L. Susskind,
\newblock ``The world as a hologram'',
\newblock J.Math.Phys.36:6377-6396,1995, hep-th/9409089. 



\bibitem{jaex}
J. Maldacena and A. Strominger,
\newblock ``${\rm AdS}_3$ black holes and a stringy exclusion principle'',
\newblock Harvard preprint HUTP-98/A016, hep-th/9804085.





\bibitem{hawkingpage}
S.W. Hawking and D. Page,
\newblock ``Thermodynamics of black holes in anti-de Sitter
space'',
\newblock \cmp\ {\bf 87} (1983) 577.


\bibitem{BTZ}
M. Banados, C. Teitelboim, and J. Zanelli,
\newblock ``The Black hole in three-dimensional space-time'',
\newblock Phys. Rev. Lett. ~{\bf 69} 1849,  hep-th/9204099 ;
M. Banados, M. Henneaux, C. Teitelboim, and J. Zanelli,
\newblock ``Geometry of the (2+1) black hole'',
\newblock \pr ~{\bf D}48 1506, gr-qc/9302012.


\bibitem{carlip}
S. Carlip,
\newblock ``The (2+1)-dimensional black hole'',
\newblock \cqg\ {\bf 12} (1995) 2853, gr-qc/9506079.


\bibitem{garydon}
G.T. Horowitz and D. Marolf,
\newblock ``A new approach to string cosmology'',
\newblock hep-th/9805207.


\bibitem{birrelldavies}
N.D. Birrell and P.C.W. Davies,
\newblock {\it Quantum fields in curved space},
\newblock Cambridge University Press (1982).




\bibitem{itzhakietal}
N. Itzhaki, J. Maldacena, J. Sonnenschein and
S. Yankielowicz,
\newblock ``Supergravity and the large-N limit
of theories with 16 supercharges'',
\newblock \pr\ {\bf D58} (1998), hep-th/9802042.


\bibitem{thermwitt}
E. Witten,
\newblock
``Anti-de Sitter space, thermal phase transition, and confinement in
gauge theories'',
\newblock
hep-th/9803131.










\end{thebibliography}
\end{document}